\documentclass[
        english,biblatex,runningheads
    ]{llncs}

\usepackage[doi=true, isbn=false, url=false]{biblatex}

\usepackage[T1]{fontenc}

\usepackage{graphicx}
\usepackage{longtable}

\usepackage{amsmath,amssymb}

\usepackage{booktabs}
\usepackage{array}
\usepackage{multirow}
\usepackage{wrapfig}
\usepackage{float}
\usepackage{colortbl}
\usepackage{pdflscape}
\usepackage{tabularx}
\usepackage[normalem]{ulem}
\usepackage{makecell}
\usepackage{multicol}
\usepackage[justification=centering]{caption}
\usepackage{todonotes}

\usepackage{tikz}

\usetikzlibrary{arrows.meta,positioning,shapes.geometric,shapes.misc, calc}

\usepackage{framed}

\providecommand{\tightlist}{%
    \setlength{\itemsep}{0pt}\setlength{\parskip}{0pt}}

\providecommand{\autocite}[1]{\cite{#1}}

\providecommand{\citep}[1]{\cite{#1}}

\begin{document}

        \title{To be FAIR or RIGHT?}
    
        \subtitle{Methodological {[}R{]}esearch {[}I{]}ntegrity
{[}G{]}iven {[}H{]}uman-facing {[}T{]}echnologies using the example of
Learning Technologies}

%\titlerunning{Abbreviated paper title}
% If the paper title is too long for the running head, you can set
% an abbreviated paper title here
%
\author{Julian Dehne\inst{1,2}\orcidID{0000-0001-9265-9619}}
\authorrunning{J. Dehne}
% First names are abbreviated in the running head.
% If there are more than two authors, 'et al.' is used.
%
\institute{German Computer Society \\
Weydingerstr. 14, 10178 Berlin, Germany \\
\email{julian.dehne@gi.de}\\
\and
University of Potsdam\\
An der Bahn 2, 14476 Potsdam, Germany \\
\email{dehne3@uni-potsdam.de}
}

    \maketitle

    \begin{abstract}
        Quality assessment of Research Software Engineering (RSE) plays
        an important role in all scientific fields. From the canonical
        three criteria, \emph{reliability}, \emph{validity}, and
        \emph{objectivity}, previous research has focussed on
        reliability and the FAIR principles. The RIGHT framework is
        introduced to fill the gap of existing frameworks for the
        validity aspect. The framework is constructed using the methods
        of theory transfer and process modelling. It is based on
        existing models of simulation research, design-based research,
        software engineering and empirical social sciences. The paper
        concludes with two case studies drawn from the field of learning
        technologies to illustrate the practical relevance of the
        framework for human-facing RSE.

    \end{abstract}

    \section{Introduction}\label{introduction}

    Research Software Engineering (RSE) is a standard process in today's
    science \autocite{brettResearchSoftwareEngineers2017a}: the term
    research software (RS) refers to software that is developed in the
    course of the research process or for a specific research purpose.
    This includes, for example, scripts used to automate data analysis
    workflows or domain-specific software libraries. General-purpose
    tools such as office software are, of course, also used in research,
    but they are not considered dedicated RS. There are some attempts to
    classify RS such as
    \autocite{hasselbringMultiDimensionalCategorizationResearch2024}.
    However, these categorizations are broader than the original
    definition that ties the software to the specific research process.
    Even though these perspectives are useful in terms of mapping the
    technical RS spectrum they contribute little to the question of
    developing a gold standard for RS development as a scientific
    activity. In fact,
    \autocite{feldererInvestigatingResearchSoftware2025b} list ``How to
    (better) organize softwarecentric scientific processes?'' as one of
    the open research questions of the newly created research area. This
    paper proposes a idealized model for the subset of human-facing RS
    of which the learning technologies community is one of the most
    prominent members.

    The problem can be illustrated with the following examples from
    different disciplines. In the geo sciences, simulation play an
    important role as some processes cannot be observed empirically due
    to their duration, complexity or unpredictability. This simulation
    of Dust Devils \autocite{tibav:9352} shows how a natural phenomenon
    can be reproduced exactly with a mathematical model. Dust Devils are
    a sand turbulence that inspired the mythical jinn figure. Due to the
    unpredictable nature of dust devils the model cannot be easily
    compared to empirical data. Consequently, the causality and
    robustness of the explanation are called into question.

    Another example is the four color problem
    \autocite{robertsonNewProofFourcolour1996} in mathematics which
    constitutes one of the first computer proofs. Here, the deviation
    from the canonical gold standard is not the lack of empirical
    analysis but the generation of knowledge instead of constructing it
    piece by piece. From the perspective of scientific scrutiny, this
    also raises questions of explainability and the intersubjective
    construction of knowledge.

    Finally, in educational technologies RS is built to serve the higher
    goal of improving learning outcomes. For instance,
    \autocite{Dehne2021} shows a complete framework both technological
    and pedagogical of supporting inquiry based learning with
    technology. In this case, the developed software changes the reality
    of teaching so much that it is hard to define what the actual
    effects are in terms of learning outcomes without including the
    software and its development process into the equation.

    The common denominator of these examples is the insight that the
    question of validity of RS is not only a technical one. In each of
    these cases, software evaluation can and should be employed to
    ensure that the code quality is high, the computations are correct
    and the software is reproducible. However, the mere openness or high
    standard of coding are not enough to answer the question, whether
    the software can be accepted as a valid epistemic instrument.

    This paper is organized as follows: First, related work regarding
    the FAIR principles and attempts at modelling research software
    engineering is discussed. Due to the interdisciplinary and
    theoretical nature of this paper, the definitions used for research,
    RS, and research software engineering are clarified. This includes a
    brief discussion of the definition of quality criteria for RS and
    how issues of validity may have received comparatively less
    attention than reliability. To tackle this research gap, and as a
    basis for the intended process modell, existing models from the
    fields of software engineering, simulation theory, design-based
    research and social sciences are discussed. Finally, the models are
    merged into a unified framework for {[}R{]}esearch {[}I{]}ntegrity
    {[}G{]}iven {[}H{]}uman-facing {[}T{]}echnologies (RIGHT-Framework).

    \section{Related Work}\label{related-work}

    The current discussion of what constitutes good RS centers around
    the question of reproducibility and the efficiency of reuse. This is
    encapsulated in the FAIR principles
    \autocite{wilkinsonFAIRGuidingPrinciples2016a}, and also represented
    on the guidelines for good RS put forward by the German Computer
    Society \autocite{czerniak2025rseguideline}. The main criteria,
    \emph{findable}, \emph{accessible}, \emph{interoperable} and
    \emph{reusable} represent the core of the open science movement.
    Despite the provocative title this paper is not challenging the
    importance or necessity of these ideals. It is critical that
    research can be replicated and reproducibility plays a main role in
    this. The research software engineering movement in turn has
    profited from the relevance and publicity and focuses on these
    aspects (see proceedings of deRSE or RSECon). In these communities
    there is intensive research of transferring software quality
    management techniques to RS engineering
    \autocite{druskatBetterArchitectureBetter2025}. Most of these
    concepts come from the STEM area. The same holds true for existing
    RSE-awards and corresponding quality criteria handed out by the
    Helmholtz research institute
    \autocite{castellQualityIndicatorResearch2024} or US-RSE
    \autocite{CommunityAwards}.

    The FAIR principles focus on quality criteria specific to the
    software. However, there are also attempts to model the research
    process with regard to computational methods:
    \autocite{jungSoftwareDevelopmentProcesses2022} studies how
    scientific software for ocean system models is developed. Using
    interviews with experts and BPMN process modeling, it describes an
    idealized development process that can help improve current
    practices in simulation-based system engineering. As an attempt to
    study the research process it is the closest attempt to this paper.
    However, as the authors focus on simulation in a natural science
    domain, a lot of aspects do not apply to the scope set here.

    Other research focuses on RS as research data and how it can be
    maintained and sustained \autocite{Anzt2021}. Outside of
    functionality and reproducibility, the environmental impact of RS's
    energy consumption has become the object of enhanced scrutiny
    \autocite{lannelongue2023}.

    \section{Definitions}\label{definitions}

    \subsection{Unit of Analysis}\label{unit-of-analysis}

    \emph{The unit of this analyse} is the subset of human-facing
    technologies, or even more specifically human-facing research
    software development. The RSE movement and the open science movement
    have their roots in the scientific computing community which in turn
    can be traced back to the natural sciences \autocite{rse2ds}. For
    this reason human-facing research such as educational technologies,
    sociological experiments, or human-computer interaction research
    have been disconnected from the discourse. Moreover, given the
    complexity and long methodological history when human bias, ethical
    concerns and such enter the research process, it can be argued that
    the RSE research has a blind spot for these fields and existing
    pipelines, models and ideals cannot be transferred without changing
    a lot of the underlying assumptions. In order to inspect the
    assumptions, we are drawing the definitions for research software
    engineering aspects not (as usual) from the software part of the
    term but look at it from the research process side. Here, we find
    that as computer scientists we can formalize the process and model
    it with existing techniques.

    \subsection{Definition of (Research)
    Software}\label{definition-of-research-software}

    Given the unit of analysis, \emph{the definition of RS} is
    restricted to mean RS that was developed specifically for a certain
    study or group of studies with planned human interaction. This is a
    specification of \autocite{gruenpeterDefiningResearchSoftware2021}.
    In contrast to traditional scientific software, there are fewer
    mathematical or other static models that would indicate that the
    software could be converted to a library. Thus, software projects in
    this areas of research tend to be more project-oriented and are less
    generalizable. In order to avoid speculation on the latter point, we
    limit the definition to RS that is incubating and has not formed a
    community of maintainers. This way, we can focus on the design /
    prototyping stage without overly restricting the term RS or
    speculating with regard to disciplinary bounderies.

    \subsection{Definition of Research
    (Software)}\label{definition-of-research-software-1}

    Research is defined more broadly: Not all research needs to be
    empirical to qualify as such. Empirical validation or falsification
    is only one paradigm. Following
    \autocite{kuhnStructureScientificRevolutions1994} we define a
    \emph{research paradigm} via these factors:

    \begin{itemize}
    \tightlist
    \item
      symbolic generalizations (scientific rules that can be formalized
      with Math or structural models)
    \item
      models: heuristic analogies or metaphors that have the function to
      visualize a domain
    \item
      values: normative and methodological assertions such as valuing
      quantifiable results higher than qualitative ones
    \item
      best practice pattern: pattern represent existing working
      solutions that function as starting points for new problems
    \end{itemize}

    \autocite{wissenschaftsrat2012} differentiates between
    \emph{empirical research} (testing, describing, experimenting),
    \emph{design-based research }(constructing, designing, innovating,
    exploring), \emph{hermeneutic research} (reconstructing,
    interpreting) and \emph{theoretical research} that consist of
    proving, arguing or systemizing (such as this paper).

    \subsection{Definition of (Research Software)
    Engineering}\label{definition-of-research-software-engineering}

    Based on the history discussed previously \autocite{rse2ds},
    research software engineering is closely connected to the empirical
    paradigm of research. Thus, research software engineering is
    connotated with programming or software development. In that sense,
    the current RSE movement should use the term Research Software
    Development to be more precise. Even though this perspective is
    valid in many cases, it disregards the disciplines that historically
    subscribed to the design-based research paradigm. In fact,
    engineering has its own academic tradition: In the 19th century,
    engineers rebelled against their status as craftsmen. As a
    consequence, technical universities were founded. The principles of
    testing and best practice designs were introduced in the academic
    system as a viable scientific method
    \autocite{lohseGrundrissWissenschaftsphilosophiePhilosophien2017}.
    Especially in Germany, the academic reputation of engineers were
    such that they were allowed supervise and award doctoral degrees
    \footnote{For this I don't have a
    reference but in the final manuscript I can reference the names}.

    Reading the history of engineering the parallels to the RSE examples
    given in the introduction become apparent. For some situations, it
    was not possible solve the need for a working solution without
    accepting the design decisions as the ground truth. This holds true
    for many RS projects. In conclusion, the term \textbf{engineering}
    fits better than expected as it includes best practices and patterns
    as the cornerstone of research paradigms. Based on this realization,
    we will include the trends of \emph{design-based research} and
    \emph{design science} into the process model.

    \subsection{Quality Criteria for Research
    (Software)}\label{quality-criteria-for-research-software}

    In the social sciences, the scientific method requires that
    empirical studies are conducted in a reliable valid, and objective
    manner \autocite{schnell1999methoden}. Social Sciences have a long
    tradition discussing these issues. For this reason, the relatively
    young fields of educational technologies and research software
    engineering can apply some of the the learnings with regard to
    human-in-the-loop aspects of human-facing technologies.

    \textbf{Reliability} expresses the minimal requirement for the
    instruments in the study: Repeated measurements should result in the
    same outcomes \autocite[132]{schnell1999methoden}. With regard to RS
    this is already captured well by the FAIR principles. But also the
    quality criteria from classical software engineering such as testing
    and fulfillment of requirements target reliability as a principle.

    \textbf{Validity} expresses the extent to which an instrument
    measures what it is supposed to measure
    \autocite[132]{schnell1999methoden}. For instance, in a survey where
    the participants are forced to answer a certain way the reliability
    is going to be very high of obtaining a certain outcome. However,
    this instrument does not inform on the real tendencies that would be
    expressed in a free study. With regard to RS the software and its
    usage need to be evaluated together. The measurement instrument can
    be defined for example as a software in the field of learning
    technologies and its corresponding usability study. The usability
    can be measured as reliably high but the supposed intent of
    measuring an increase or decrease in learning outcomes is not
    achieved.

    \textbf{Criteria Validity} is achieved in two ways. \emph{predictive
    validity} and \emph{concurrent validity}
    \autocite[137]{schnell1999methoden}. In both cases external
    independent criteria are used to compare the expected result and the
    prediction. In the case of concurrent validity, known groups are
    used where the difference between the known groups are a function.
    This is similar to metamorphic testing
    \autocite{chen1998metamorphic} in the software engineering field.
    However, as the selection of the criteria and known groups are
    difficult, this type of validity, even though used commonly, is
    regarded as problematic and linked more to reliability studies.

    \textbf{Construct Validity} is defined as a list of requirements
    \autocite[139]{schnell1999methoden}:

    \begin{enumerate}
    \def\labelenumi{\arabic{enumi}.}
    \tightlist
    \item
      the theoretical relationships between the constructs (RS and
      empirical study) need to be explained
    \item
      the empirical relationship between the constructs need to be
      existent
    \item
      found relationships need to be interpretable
    \item
      a newly introduced construct needs to be able to show new results
    \end{enumerate}

    This paper deals mainly with construct validity by answering the
    theoretical question (1) and providing a framework and guidelines
    for 2-4. Due to constraints of time and existing literature, the
    third quality aspect (objectivity) is outside the scope of this
    paper whereas reliability is already dealt with in the mainstream
    research.

    \section{Related Process Models}\label{related-process-models}

    Following the previous definition of research paradigms, we assume
    that models and best practices can be transferred between
    disciplines and reconstructed as heuristic analogy
    \autocite{kornmesserWissenschaftstheorieEinfuehrung2020} for a new
    field of study (theory transfer). In order to derive an ideal
    process model for human-facing research software engineering we draw
    on existing models from \emph{simulation theory}, \emph{design-based
    research}, \emph{software engineering} and \emph{empirical social
    sciences}. These input theories are predetermined by the scope of
    this paper in terms of social sciences and software engineering.
    From the definition of engineering above the design-based research
    model was implied. The simulation model was included because it
    showcases how research software engineering transforms the
    scientific practice. Even though it is more relevant for
    STEM-disciplines, it also plays a role in modern computational
    social sciences such as complexity theory.

    The process models are not depicted using classical modelling
    languages from software engineering such as UML or BPMN. For one,
    these languages do not capture the abstract nature of the models.
    Moreover, a domain-agnostic language was needed. For this reason,
    the graphs can be interpreted as directed acyclic graphs (DAGs) from
    statistics \autocite{10.1001/jama.2022.1816}. This also fits the
    methodological frame of this paper. The nodes either represent data
    sets of a certain stage or boolean decision variables in the case of
    decision points.

    \subsection{Simulation}\label{simulation}

    As shown in the example of the dust devils, simulation as a method
    becomes the only choice for certain phenomena and research
    questions. The reason, why this methodology is considered powerful
    is that it has a high built-in external validity and construct
    validity: on the one hand a classical empirical study is conducted,
    on the other hand a theoretical model is used to find so-called
    mechanisms that can predict empirical patterns mathematically.

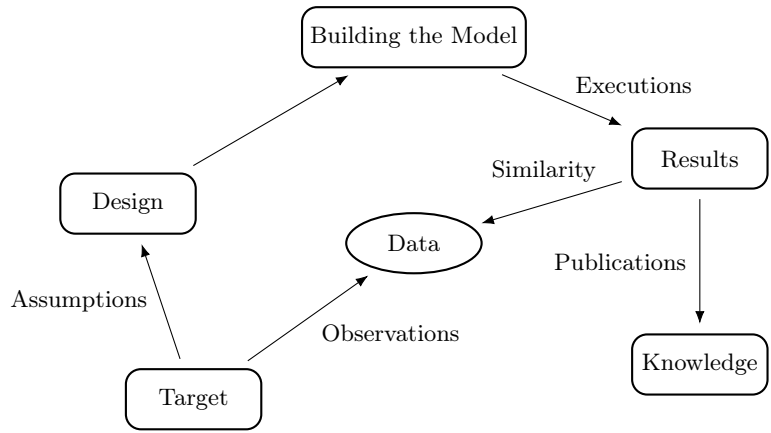
\begin{figure}[h!]
    \centering
    \begin{tikzpicture}[
        node distance=18mm,
        >=Latex,
        every node/.style={outer sep=2pt},
        every path/.style={shorten >=2pt, shorten <=2pt},
        box/.style={draw,
        rounded corners=2mm,
        thick,
        minimum width=18mm,
        minimum height=8mm,
        align=center}
    ]
        \node[box](build) {Building the Model};
        \node[box, below left=of build] (design) {Design};
        \node[box, below right=of build, yshift=6mm] (results) {Results};
        \node[box, ellipse, below=of build] (data) {Data};
        \node[box, below left=of data] (target) {Target};
        \node[box, below=of results] (knowledge) {Knowledge};

        \draw[->] (build) -- node[midway, right, yshift=2mm]{Executions} (results);
        \draw[->] (design) -- (build);
        \draw[->] (target) -- node[midway, left]{Assumptions} (design);
        \draw[->] (target) -- node[midway, right, yshift=-2mm]{Observations} (data);
        \draw[->] (results) -- node[midway, above, yshift=1mm, xshift=-1mm]{Similarity} (data);
        \draw[->] (results) -- node[midway, left]{Publications} (knowledge);

    \end{tikzpicture}
    \caption{The logic of simulation as a method. Diagram designed by \autocite{drogoulMultiagentBasedSimulation2003}}
    \label{fig:simulation}
\end{figure}

    Figure \ref{fig:simulation} shows the generalized model from
    \autocite{drogoulMultiagentBasedSimulation2003}. First, the
    assumptions are derived from the target model. Based on these
    assumptions certain design-decisions are taken. These encoded
    mechanisms form the model. Based on the model, synthetic results are
    produced during the simulation stage. Concurrently, observations are
    made based on the target model that produce empirical data. A
    simulation study is deemed valid if and only if the empirical data
    matches the synthetic data to a high extend.

    In the realm of human-facing technologies these studies are rare but
    a promising way of stabilizing research outcomes: for instance, an
    agent based model of peer assessment could be constructed. The
    simulated students give each other grades. This could then be
    contrasted to a case study where peer assessment was tried in the
    real world.

    \subsection{Design-based Research}\label{design-based-research}

    There are two similar approaches fromt the domains of educational
    innovation and applied computer science: Design-Based Research (DBR)
    is frequently employed in educational and medical contexts, where
    the focus lies on interventions aimed at improving learning or
    practice. In contrast, Design Science Research (DSR) primarily
    emphasizes the development and evaluation of technological
    artifacts.

    Methodologically, DBR typically proceeds through numerous iterative
    cycles, in which researchers progressively refine interventions and
    move step by step toward an improved solution for a given
    (educational) problem \autocite{reinmann2005innovation}. DSR, by
    comparison, is often conceptualized as involving multiple parallel
    research cycles \autocite{vombrockeIntroductionDesignScience2020}.
    These cycles simultaneously contribute to strengthening the
    theoretical foundation, advancing the solution to the practical
    problem, and refining the underlying design principles. As a result,
    the DSR process is often longer and more complex. In DBR,
    functioning prototypes frequently emerge throughout the iterative
    process, whereas in DSR fully developed artifacts commonly appear
    only toward the end of the research process.

    \begin{figure}[h!]
    \centering
% Preamble:
% \usepackage{tikz}
% \usetikzlibrary{arrows.meta,positioning,calc}

% Preamble
% \usepackage{tikz}
% \usetikzlibrary{arrows.meta,positioning,calc}

    \begin{tikzpicture}[font=\sffamily]

% --- Colors
        \definecolor{ringblue}{RGB}{55,98,150}
        \definecolor{arrowyellow}{RGB}{255,204,0}

% --- Styles
        \tikzset{
            labelbox/.style={
                draw=ringblue,
                line width=1pt,
                fill=white,
                rounded corners=2pt,
                inner xsep=8pt,
                inner ysep=6pt,
                align=center
            },
            cyclearrow/.style={
                -{Stealth[length=6mm,width=6mm]},
                line width=6pt,
                draw=arrowyellow
            }
        }

% --- Geometry (ADJUST HERE)
        \def\R{2.2}        % smaller radius
        \def\ringW{20pt}   % thicker ring

% --- Thick ring
        \draw[draw=ringblue,line width=\ringW] (0,0) circle (\R);

% --- Arrows on ring
        \draw[cyclearrow] (150:\R) arc[start angle=150,end angle=100,radius=\R];
        \draw[cyclearrow] (25:\R)  arc[start angle=25,end angle=-35,radius=\R];
        \draw[cyclearrow] (-95:\R) arc[start angle=-95,end angle=-155,radius=\R];

% --- Outer labels
        \node[labelbox] at (0,\R+1.0) {Analysis /\\ Exploration};
        \node[labelbox] at (\R+1.8,0.1) {Design /\\ Construction};
        \node[labelbox] at (-\R-1.6,-0.6) {Evaluation /\\ Reflexion};

        \node[labelbox] at (0,0) {Testing/\\ Improvement};

% --- Bottom-left evaluation maturity node
        \node[labelbox, text width=5.2cm] (maturity)
        at (-\R-2.2,-\R-2.4)
            {The intervention has reached a level of maturity\\
        that allows it to be evaluated summatively};

% --- Arrow pointing TO that node
        \draw[cyclearrow, shorten >=5mm] (-\R,-\R) -- (maturity.north);

    \end{tikzpicture}

    \caption{The logic of design-based research as a method. Diagram and process model adapted from \autocite{reinmann2005innovation}}
    \label{fig:dbr}

\end{figure}
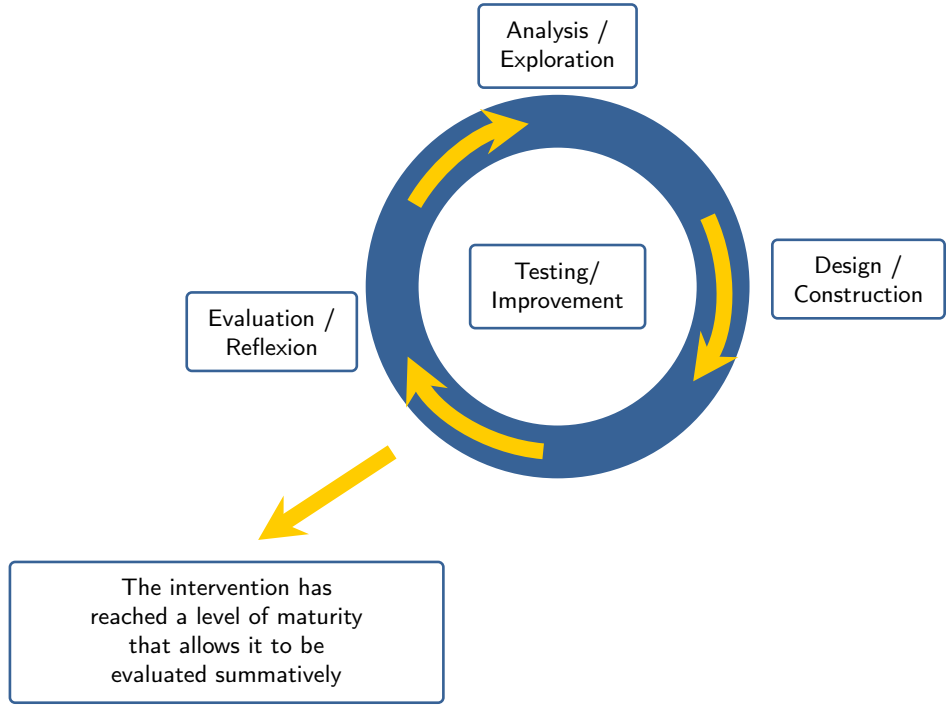

    A common denominator of both approaches is the focus on design
    principles and a continuous refinement of the latter. Another
    commonality is that the process does not merely follow the
    requirements (as they are virtual) but also include a vision of the
    future. In DBR this is called ``what could be''. In DSR it is
    formulated as the ``motivation''. As our scope is human-facing
    technologies, the DBR approach seems closer to the point as it
    includes a semi-agile idea of a high turnover in educational
    interventions to gather information of actual human use (see figure
    \ref{fig:dbr}). In order to keep the merged process modell readable,
    we will use the above simplified modell of DBR. It should be noted,
    that there is no inherent conceptual conflict between the approaches
    and for the application case of human-facing technologies both could
    be merged to include an elaborate technical design-stage and a
    human-intervention within a bigger loop.

    \subsection{Software Engineering}\label{software-engineering}

    Another perspective on modelling research software engineering is to
    regard it as specification of software engineering. Here, there
    exist a number of widely recognized process models such as
    \autocite{balzertLehrbuchSoftwaretechnikBasiskonzepte2010a} or
    \autocite{sommervilleSoftwareEngineering2002}. Due to their
    visibility, the waterfall model and the requirement-based software
    engineering model of
    \autocite{balzertLehrbuchSoftwaretechnikBasiskonzepte2010a} are not
    depicted here. The question remains, if these could be used as basis
    for our intended model.

    \begin{table}[!h]
    \caption{Comparison of software engineering approaches. R denotes Requirements. U denotes User. V denotes Validity.}
    \label{tbl:se}

    \begin{tabular}{l|l|l|l|l}

        & \textbf{SE} & \textbf{Agile SE} & \textbf{Scientific SE} & \textbf{Design-based SE} \\
        \hline

        \textbf{R} & R are clear
        & R change
        & Emerging during work
        & Intentions instead of R \\

        \textbf{U} & U are known
        & U are involved
        & U are developer
        & U are hypothetical \\

        \textbf{V} & \makecell[tl]{R tracing \\ and Testing}
        & Continuous Feedback
        & \makecell[tl]{Experiments or  \\ Mathematical proofs}
        & \makecell[tl]{Best practices \\ and positive evaluation} \\

    \bottomrule

    \end{tabular}

\end{table}

    We argue that even though these models form the basis of any
    software development, they are incomplete with regard to
    human-facing research software development. Table \ref{tbl:se}
    illustrates the difference on three criteria: how are requirements
    defined and how clear are they? (R), what is the concept of a user
    and how much information is there? (U) and how is evaluation
    conceived and connected to validity? (V).

    In traditional software engineering (SE), requirements are assumed
    to be clear and stable, users are known in advance, and validity is
    ensured through requirements tracing and systematic testing. In
    agile software engineering, requirements change during development.
    Users are actively involved in the process, and validity is achieved
    through continuous feedback and iterative improvement
    \autocite{fowler2001agile}. In scientific software development (as
    defined earlier for STEM-disciplines), requirements emerge during
    the work, and the developer is often the main user. Validity is
    established through experiments, tests or mathematical proofs. In
    design-based software engineering, development is guided by
    intentions rather than fixed requirements, users are often
    hypothetical, and validity relies on best practices and positive
    evaluation.

    In summary, classical software engineering models form the basis for
    the human-facing research software engineering, too. However, they
    lack coverage with regard to the specific validity constraints and
    user constellations in the human-facing research domains.

    \subsection{Empirical Research (Social
    Sciences)}\label{empirical-research-social-sciences}

\begin{figure}[h!]
    \centering
    \newcommand{\rsecycle}{\includegraphics[height=12mm]{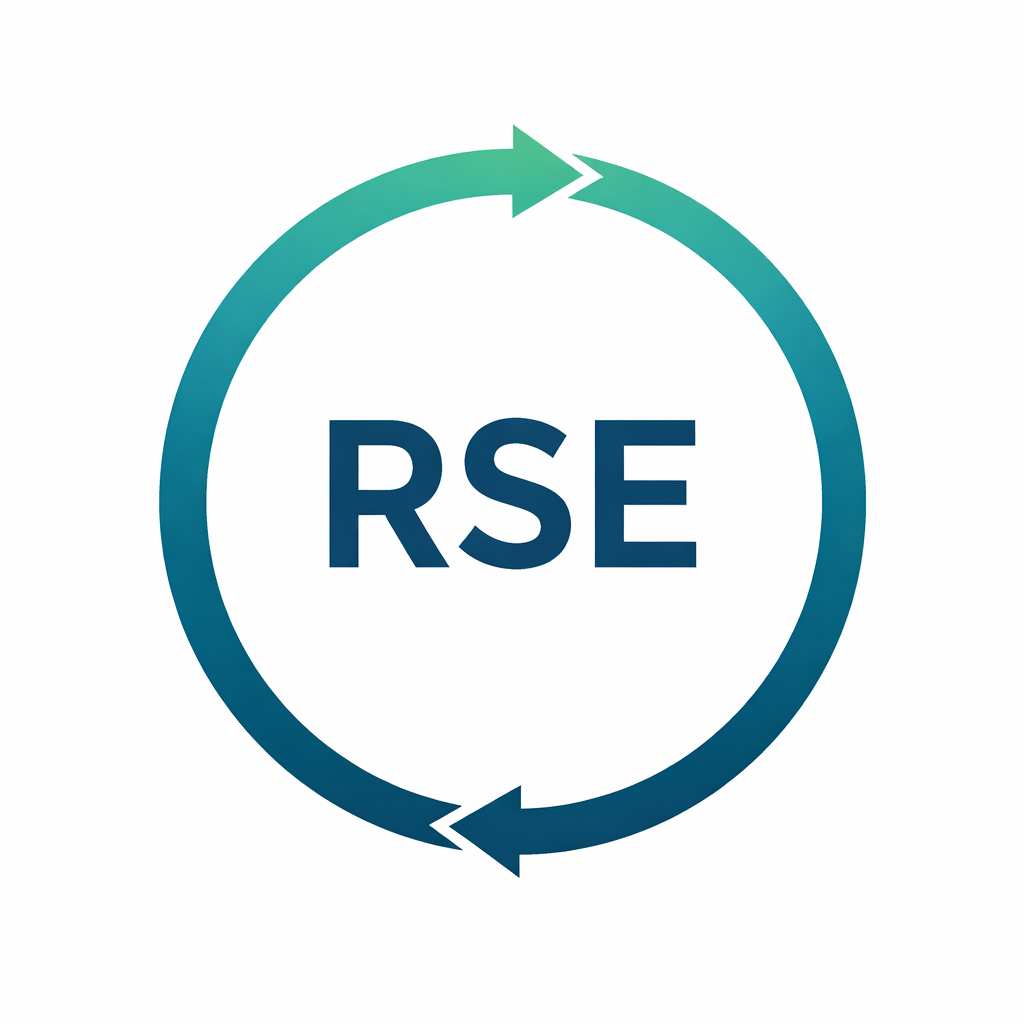}}

    \tikzset{
        box/.style={draw,
        rounded corners=2mm,
        thick,
        minimum width=18mm,
        minimum height=8mm,
        align=center},
        rsepic/.style={
            inner sep=0pt,
            outer sep=2pt,
        }
    }

    \begin{tikzpicture}[
        node distance=8mm,
        >=Latex,
        every node/.style={outer sep=2pt},
        every path/.style={shorten >=2pt, shorten <=2pt}
    ]

% --- main vertical chain ---
        \node[box] (problem) {Selection of\\Research Problem};
        \node[box, below=of problem] (theory) {Theory Building};

% --- side boxes (hatched) ---
        \node[box, below left=of theory] (concept) {Concept Specification\\Operationalization};
        \node[box, below right=of theory] (form) {Determination of\\Study Design};

% --- right side (hatched) ---
        \node[box, below=of form] (unit) {Selection of\\Unit of Analysis};
        \node[box,  below=of unit] (collect) {Data Collection};
        \node[rsepic, right=of collect, xshift=-8mm] (rse_collect) {\rsecycle};

        \node[box,  below=of collect] (record) {Data Recording};
        \node[rsepic, right=of record, xshift=-8mm] (rse_record) {\rsecycle};
        \node[box,  below=of record] (analysis) {Data Analysis};
        \node[rsepic, right=of analysis, xshift=-8mm] (rse_analysis) {\rsecycle};

% --- left side (hatched) ---
% \node[box, below right=of concept, xshift=-8mm] (thought) {Thought Experiment};
        \node[box, below=of concept] (simulation) {Simulation};
        \node[rsepic, left=of simulation, xshift=+8mm] (rse_simulation) {\rsecycle};

        \node[box,  below left=of analysis] (comparison) {Similarity Assessment \\ Synthesis};
        \node[box,  below=of comparison] (pub) {Publication};

        \draw[->](problem.south) -- (theory.north);

% --- connectors from theory to side boxes ---
        \draw[->](theory.west) -| (concept.north);
        \draw[->](theory.east) -| (form.north);

% --- connectors from side boxes down to "unit selection" ---
        \draw[->](form.south)    -- (unit.north);
        \draw[->] (unit.south)    -- (collect.north);
        \draw[->] (collect.south)    -- (record.north);
        \draw[->] (record.south)    -- (analysis.north);
        \draw[->] (analysis.south)    |- (comparison.east);

% --- connectors from side boxes down to "simulation" ---
        \draw[->] (concept.south)    -- (simulation.north);
%\draw (concept.east)    |- (unit.west);

        \draw[dotted,<->] (concept.east) -- node[midway, below left]{influences}  (unit.west);

        \draw[->] (simulation.south)    |- (comparison.west);

        \draw[->] (comparison.south)    -- (pub.north);

    \end{tikzpicture}

    \caption{The research process in the empirical social sciences based on \cite{schnell1999methoden} }
    \label{fig:schnell}

\end{figure}
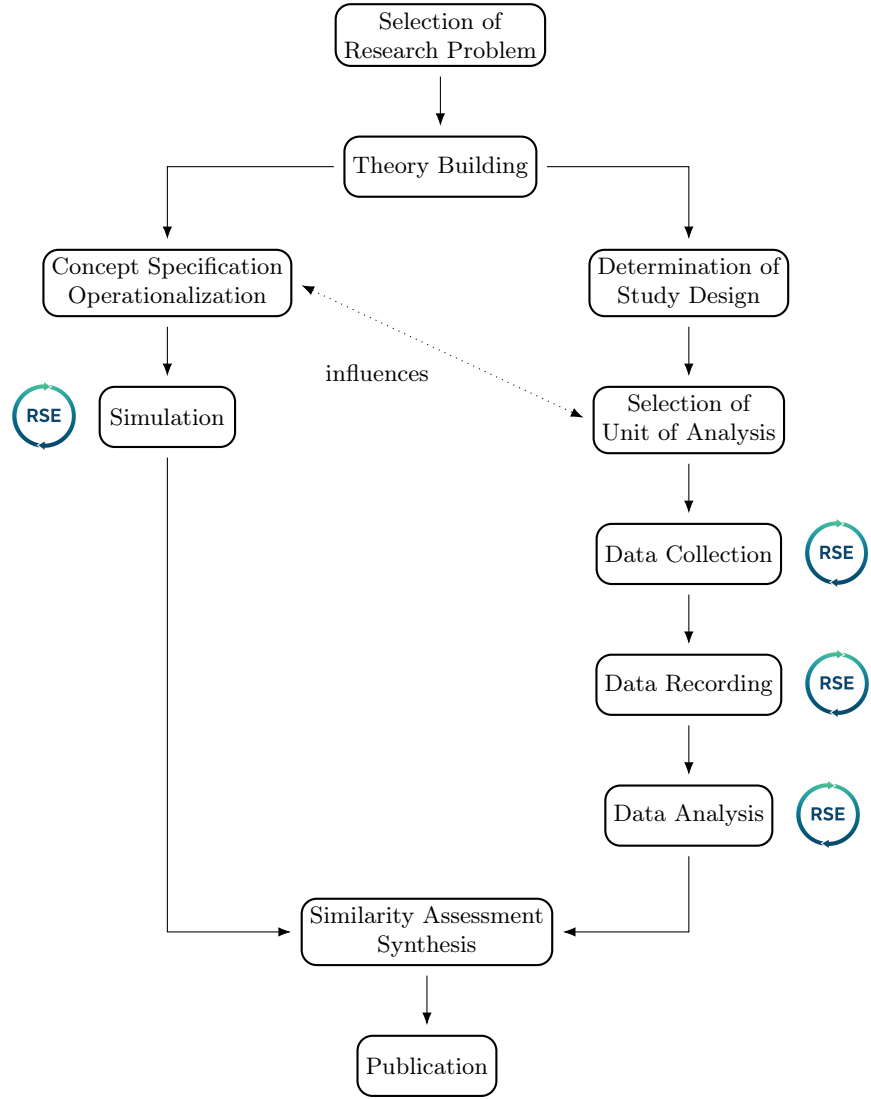

    As the last input model, we introduce the research process as it is
    defined in the empirical social sciences. Figure \ref{fig:schnell}
    shows the different stages. As discussed earlier we are using DAG
    notation to model the process. This means that nodes can be the
    perfect correlation function which in practice means that they can
    be stepped over.

    Based on the research process by \autocite{schnell1999methoden},
    figure \ref{fig:schnell} was adapted to include both the simulation
    concept and a label where RSE may play a role. The process model in
    the empirical social sciences centers around the process of making
    the theoretical assumptions measurable: After connecting the study
    to a school of thought, the study is designed focussing on two
    aspects: as a first step, the unit of analysis is determined. Given
    the interconnectedness of the social world, it is exceedingly
    difficult to isolate variable for closer inspection. This process is
    influenced by theoretical work on how a concept can be
    operationalization which means that the variables are broken down
    into measurable items and assumed connections. These assumed
    connections feed into the \emph{mechanisms} that underlie the
    simulation. If the simulation method is chosen, RSE is required to
    execute this form of research. If not, then the assumptions of the
    operationalization are compared to the results of the empirical
    study directly.

    As a second step, the empirical study is executed including the
    classical process of data collection, recording and analysis. In
    each of these stages, complex software development process may take
    place. As a leading example you can think of social media study
    where the response trees are modelled and simulated and then
    compared to empirical data from a social platform that in turn needs
    to be collected, stored and analyzed using computational methods. It
    should be noted that each of these partial steps can contain a
    bigger RSE project. We would like to draw attention to certain
    characteristics of this model: the requirements for the research are
    a moving goalpost and technical issues with the software may only be
    detected at a later stage in the research process which may be
    detrimental to the research project. For this reason, the
    reliability (and the FAIR principles) play a critical role in
    ensuring that the study can proceed despite the high number of
    potential points of failure. This aspect is exacerbated if you
    consider the potential that some of the RSE projects may cover
    uncharted ground which implies a design-based inner loop. This
    illustrates the problem of exchanging the traditional methods with
    computational ones without reflecting the research process as a
    whole.

    Another aspect is the fact, that each of these steps has different
    contexts and thus very different outcomes in terms of RS. In order
    to systematize the best practices in more details, a detailed
    classification of RS in the social sciences based on the function in
    the research process is needed.

    This process modell already covers the mainstream research in both
    the social sciences and the educational sciences given that the
    empirical paradigm is dominant. For simplicity of the argument we
    assert that educational science counts as social sciences if the
    empirical paradigm is prevalent. In the following, we argue that
    even though the human factor is coded into the definition of social
    sciences, the process model does not the cover the scope of
    human-facing technologies.

    For one, even today not all social sciences use RS. But even if the
    objects of the analysis are human made, the RS that is used for
    analysis does not necessarily face humans other than researchers. In
    this argument we exclude the researcher bias as special case.
    Earlier we stated that objectivity is outside the scope of this
    investigation. Thus, even though the researchers might be biased
    concerning human data because of gender, race or political leanings,
    this is more a question of objectivity rather than validity. For
    these reasons, the empirical process modell should be considered a
    relevant part of the final model but needs to be specified to
    account for the special case of having \textbf{humans that are both
    users of the RS and in some sort of relationship to the variables of
    the research as such}.

    \section{A Model for Research Software Development with
    Humans-In-The-Loop}\label{a-model-for-research-software-development-with-humans-in-the-loop}

    RS is often discussed as part of the empirical research pipeline,
    but its methodological role differs depending on how it enters the
    study. The key point in this paper is that RS can (i) influence
    relationships among variables while leaving the empirical variable
    universe essentially unchanged, or (ii) transform the research
    context in such a way that the study becomes possible only with RS,
    because RS changes what can be treated as an input (independent
    variable), what can be measured or optimized, and even what counts
    as an outcome (dependent variable). We formalize this distinction
    with a lightweight context--variable model and connect it directly
    to causal diagrams.

    Let $\mathcal U$ denote the universe of potentially relevant empirical variables in a domain (e.g., learner traits, learning behaviors, outcomes, interaction measures). A concrete empirical study is conducted in a research context $C$ (e.g., classroom size, available infrastructure, permitted data sources, institutional constraints). The context determines which variables are practically observable and meaningful. We denote this context-dependent set of variables by
\[
    \mathcal V(C) \subseteq \mathcal U .
\]
To connect to common empirical practice, we separate the variables in $\mathcal V(C)$ by their role:
\[
    \mathcal V(C)= I(C)\ \cup\ D(C)\ \cup\ Z(C)
\]
With this notation we can describe two conceptually different roles of research software.

\subsection{Case 1: Research software influences relationships between variables.}\label{subsec:case-1:-research-software-influences-relationships-between-variables.}

In this situation the software affects behavior, measurement quality, or the strength of statistical or causal relationships, while the variable roles remain stable in the sense that the study can still be posed and executed without the software. Formally, for a baseline alternative $RS_0$ (for example a paper-based procedure) the study remains feasible and the independent and dependent variables remain the same except for the addition of RS as a mediator variable.
The software may nevertheless influence the empirical relationships between variables, for example by changing how students practice or how often feedback occurs.

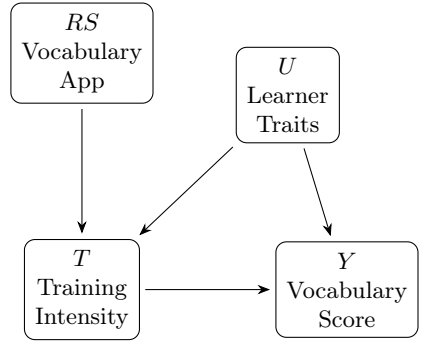
\begin{figure}[h!]
    \centering

    \begin{tikzpicture}[>=Stealth, node distance=18mm, every path/.style={shorten >=2pt, shorten <=2pt}]

        \tikzstyle{node}=[draw, rounded corners, align=center, inner sep=4pt]

        \node[node] (RS) {$RS$\\Vocabulary\\App};
        \node[node, below=of RS] (T) {$T$\\Training\\Intensity};
        \node[node, right=of T] (Y) {$Y$\\Vocabulary\\Score};
        \node[node, above right=of T] (U) {$U$\\Learner\\Traits};

        \draw[->] (RS) -- (T);
        \draw[->] (T) -- (Y);
        \draw[->] (U) -- (T);
        \draw[->] (U) -- (Y);

    \end{tikzpicture}

    \caption{Case 1 (App). Variables: $RS$ (app), $U$ (traits), $T$ (training), $Y$ (outcome).}
    \label{fig_case1}
\end{figure}

A concrete example from computer-enhanced learning is vocabulary memorization supported by a learning app. Let $U$ denote learner traits such as prior knowledge or motivation, let $T$ denote training intensity (for example practice frequency or time-on-task), and let $Y$ denote learning outcome such as a vocabulary test score. Whether training is performed using an app ($RS$) or a traditional textbook ($RS_0$), the empirical question \textsl{How does training affect vocabulary learning?} remains meaningful. The software influences the pathway from training to learning outcomes, but it does not introduce fundamentally new variables. This situation is illustrated in Figure~\ref{fig_case1}. In this situation, RS is added to the set of variables:

\[
    \mathcal D(C)=  D(C)\ \cup\ \{RS\}
\]

\subsection{Case 2: Research software transforms the research context.}\label{subsec:case-2:-research-software-transforms-the-research-context.}

In the second situation the introduction of research software changes the research context itself and thereby changes the set of variables that are available or meaningful. Formally,
\[
    \mathcal V_{RS}(C)\neq \mathcal V(C).
\]
This typically implies that the roles of variables also change:
\[
    I_{RS}(C)\neq I(C)
    \quad \text{or} \quad
    D_{RS}(C)\neq D(C).
\]
In many cases the study is only feasible when the software is available:
\[
    \phi(RQ,C,RS)=1,
    \qquad
    \phi(RQ,C,RS_0)=0.
\]
An example from computer-enhanced learning is algorithmic group formation in large classrooms. Suppose an instructor wants to study the influence of group composition on collaborative learning outcomes. In a class with a very large number of students, assigning groups according to complex constraints or optimization criteria may be practically impossible without software support. Introducing a group formation algorithm enables such assignments and therefore changes the research context.

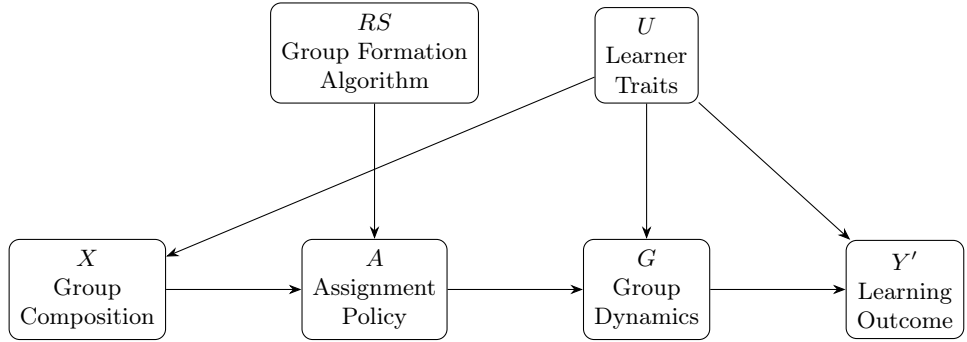
\begin{figure}[ht!]
    \centering

    \begin{tikzpicture}[>=Stealth, node distance=18mm]

        \tikzstyle{node}=[draw, rounded corners, align=center, inner sep=4pt, every path/.style={shorten >=2pt, shorten <=2pt}]

        \node[node] (RS) {$RS$\\Group Formation\\Algorithm};
        \node[node, below=of RS] (A) {$A$\\Assignment\\Policy};
        \node[node, left=of A] (X) {$X$\\Group\\Composition};
        \node[node, right=of A] (G) {$G$\\Group\\Dynamics};
        \node[node, right=of G] (Yp) {$Y'$\\Learning\\Outcome};
        \node[node, above=of G] (U) {$U$\\Learner\\Traits};

        \draw[->] (RS) -- (A);
        \draw[->] (X) -- (A);
        \draw[->] (A) -- (G);
        \draw[->] (G) -- (Yp);
        \draw[->] (U) -- (G);
        \draw[->] (U) -- (Yp);
        \draw[->] (U) -- (X);

    \end{tikzpicture}
    \caption{Case 2 (Grouping). Variables: $RS$ (algorithm), $X$ (composition), $A$ (assignment), $G$ (dynamics), $Y'$ (outcome), $U$ (traits).}
    \label{fig_case2}

\end{figure}

With the algorithm, new independent variables may appear, such as the assignment policy $A$ produced by the algorithm and the composition variables $X$ describing group structure. At the same time, new dependent variables may become relevant, such as group dynamics $G$ or collaborative learning outcomes $Y'$. These variables are closely tied to the presence of the software system that organizes and records the group interaction. This situation is illustrated in Figure~\ref{fig_case2}. The conceptual difference between the two cases can therefore be summarized informally as follows. In the first case, research software changes how an already definable study is executed. In the second case, research software changes what study can be defined and carried out at all, because it transforms the research context and thereby changes the variables that enter the empirical model.

There is a critical difference in this case in comparison to the first. Not only the variables change but also their interpretation. Going back to the empricical model of research this could be considered a loop back to the theory and operationalization stages. In this example, learning outcomes of group learning are not the same as the aggregation of individual learning outcomes. It may well be that the group as social artifact now has gained some skills that the individuals could not reproduce independently in a different context. This concept of group learning has major implications on the interpretation of the variable learning outcome. In this case, a new iteration of the design phase is needed to adjust the study to the new context changing the methodological basis from an empirical grounded approach to DBR.

    \subsection{The RIGHT decisions}\label{the-right-decisions}

\begin{figure}[h!]
    \centering

    \begin{tikzpicture}[
        node distance=8mm,
        >=Latex,
        every node/.style={outer sep=2pt},
        every path/.style={shorten >=2pt, shorten <=2pt},
        box/.style={draw,
        rounded corners=2mm,
        thick,
        minimum width=18mm,
        minimum height=8mm,
        align=center},
        rsepic/.style={
            inner sep=0pt,
            outer sep=2pt,
        }
    ]

% --- left vertical chain ---
        \node[box] (isRSE) {D1: Research Software relevant?};
        \node[box, below=of isRSE] (isHuman) {D2: Human-In-The-Loop?};
        \node[box, below=of isHuman] (interactsWithRQ) {D3: Interacts with Research Question?};
        \node[box, below=of interactsWithRQ] (transformsRQ) {D4: Transforms Research Context?};

% --- right vertical chain ---
        \node[box, right=of isRSE] (conductStudy) {S1: Conduct Empirical Study};
        \node[box, below=of conductStudy] (evaluateSoftware) {M4: Evaluate Software};
        \node[box, below=of evaluateSoftware] (evaluateUsability) {M3: Evaluate Usability};
        \node[box, below=of evaluateUsability] (softwareEffectStudy) {M2: Wizard-of-Oz Study};

% --- left vertical arrows ---
        \draw[->] (isRSE.east) -- node[midway, above]{no} (conductStudy.west);
        \draw[->] (isHuman.east) -- node[midway, above]{no} (evaluateSoftware.west);
        \draw[->] (interactsWithRQ.east) -- node[midway, above]{no} (evaluateUsability.west);

        \draw[->] (isRSE.south) -- node[midway, left]{yes} (isHuman.north);
        \draw[->] (isHuman.south) -- node[midway, left]{yes} (interactsWithRQ.north);
        \draw[->] (interactsWithRQ.south) -- node[midway, left]{yes} (transformsRQ.north);

% --- right vertical arrows ---
        \draw[<-] (conductStudy.south) -- (evaluateSoftware.north);
        \draw[<-] (evaluateSoftware.south) -- (evaluateUsability.north);
        \draw[<-] (evaluateUsability.south) -- (softwareEffectStudy.north);

% final stage
        \node[box, below=of transformsRQ] (dbr) {S2: Design-Based Research};
        \node[box, below=of softwareEffectStudy] (noStimulusControl) {M1: No-Stimulus Control};

        \draw[->] (transformsRQ.south) -- node[midway, left]{yes} (dbr.north);
        \draw[->] (transformsRQ.east) -- node[midway, below left]{no} (noStimulusControl.west);

        \draw[->] (noStimulusControl.north) -- (softwareEffectStudy.south);
        \draw[->] (noStimulusControl.east) -- ++(1,0) -- ++(0,3.5) -- (evaluateUsability.east);

%\draw[->] (A) -- ++(1,0) -- ++(0,2) -- (B)%;

    \end{tikzpicture}

    \caption{An idealized decision model for transformative empirical research software}
    \label{fig:goldstandard}

\end{figure}
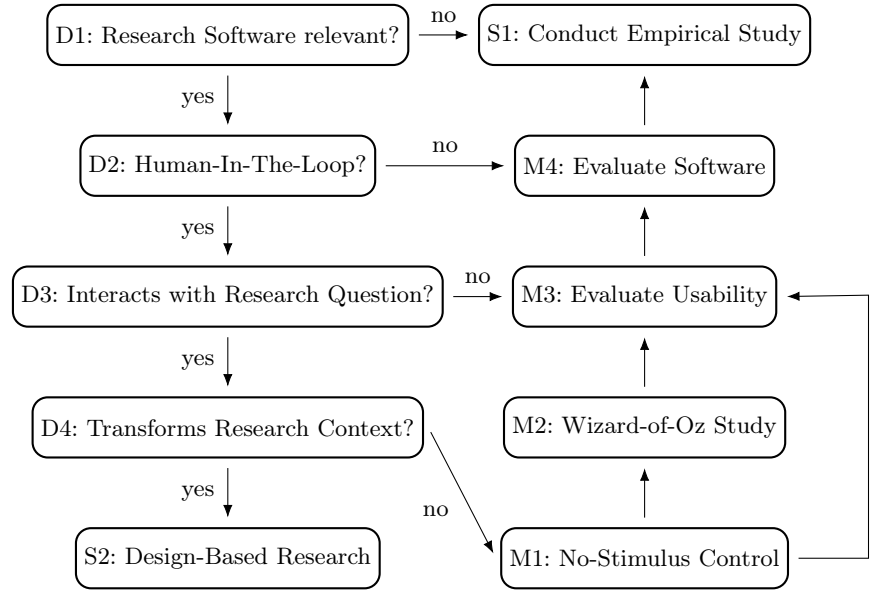

    Figure \ref{fig:goldstandard} integrates the different process
    modells into a metamodell. It highlights important methodological
    decision points that could help researchers decide which process to
    follow in order to avoid problems with validity at a later point. On
    the left side the decision points are labeled D1 to D4. On the right
    side, matching empirical methods are suggested to test for construct
    validity. As demonstrated in the use cases above, the edge cases are
    situations where the whole research paradigm may shift.

    \paragraph{D1}

    The first essential question is whether the software is relevant for
    the research. Surprisingly, this is not a trivial case.
    Technological trends such as the LLM-hype may lead to forcing the
    use of technology even if there are known alternatives with less
    obfuscating effects. Alternatively, feelings ownership or path
    dependency may encourage researchers to develop a RS independent of
    its necessity for knowledge gain. If the answer to this question is
    no, the empirical study can proceed normally. If this is the case,
    D2 follows.

    \paragraph {D2}

    For the scope of this paper the next relevant question is whether
    there are humans-in-the-loop that could lead to added complexity in
    terms of bias, added variables or interaction effects with the users
    also being part of the unit of analysis. If this is not the case,
    the classical software evaluation process should be used. This
    including testing, requirement tracing or potential formal methods
    of software verification (M4). If this is the case, D3 follows.

    \paragraph {D3}

    As shown in the case study 1, the question D2 should be specified to
    ask whether the humans facing the technology interact with the
    research question. If this is not the case, it should be tested that
    the usability of the software (M3) does not cause organizational
    problems such as dropouts or biased participants. If an interaction
    effect can be postulated, D4 follows.

    \paragraph {D4}

    As illustrated in case study 2, the question D2 can be specified to
    ask whether the RS changes the research context both in the set of
    variables and the interpretation of the variables. If this is not
    the case, the postulated interaction of D3 should be controlled
    concurrent validity (one group with the RS stimulus and one without:
    M1). As it is often not possible to recreate the experimental
    situation due to ethical or logistical reasons, the Wizard-of-Oz
    study design (M2) can be used. This means, that the effect of the RS
    is replicated without its actual use. For instance, a chatbot may be
    used which in fact has humans write the answers or some
    computational results (formed groups or similar) may be presented as
    the result of an algorithm whilst being random. If indeed, the
    research context is transformed by the RS than a design-based
    research approach should be considered.

    The more complex the research becomes due to the effects of the RS,
    the more methods need to be used to ensure the continued validity.
    This means that M1 in most cases needs to be followed by M2-M4. This
    is similar to regression testing where the simple edge cases need to
    be checked continuously.

    \section{Limitations and
    Discussion}\label{limitations-and-discussion}

    From the analysis of the process modells of simulations,
    design-based research, classical software engineering and empirical
    research a meta decision model was synthesized that helps ensure
    construct validity in the case of complexer szenarios of
    human-facing technologies. The abbreviation RIGHT does not only
    stand for the scope of the investigation ({[}R{]}esearch
    {[}I{]}ntegrity {[}G{]}iven {[}H{]}uman-facing {[}T{]}echnologies)
    but also can be used as a mnemonic for the important decision
    points:

    \begin{itemize}
    \tightlist
    \item
      {[}R{]}elevance: Is this specific RS really necessary to conduct
      the research?
    \item
      {[}I{]}nteraction: Does the RS interact with variables during the
      execution phase?
    \item
      {[}G{]}eneration: Can the empirical results be synthetically
      generated to allow for the simulation method?
    \item
      {[}H{]}uman-In-the-Loop: Are humans both users of the RS and part
      of the research context?
    \item
      {[}T{]}ransformation: Does the RS transform the research context
      and forces reinterpretation of the variables?
    \end{itemize}

    This framework comes with certain limitations: it is only
    theoretically grounded for the intersection of empirical studies,
    RSE and design-based (educational) research. Even though some of the
    ideas may be generalizable for RSE in other disciplines, its
    usefulness may be limited. Moreover, this analysis is theoretical
    only. A meta-science study that reviews existing publications in the
    ECTEL community based on the RIGHT-framework was submitted by
    {[}anonymized{]}. From this, usefulness for this framework can be
    implied but not proven. A survey study targeting the DELFI community
    is also under way and could support or disprove some of the
    assumptions of this study. Hopefully, this will add external
    validity to this study, too.

    Both the process models and the definitions could be formulated and
    modelled in a more mathematical fashion. However, due to the
    interdisciplinary nature of the input models and the readers, a more
    text-centric approach was taken.

    This analysis leads to another formulation of the educational/social
    innovation dilemma: the more transformative and socially motivated
    an innovation is, the more rigorous the evaluation needs to be to
    ensure scientific validity. It also endangers its success as a
    research output as the design-based research paradigm is less
    supported in publication outlets than empirical studies.

    This study fills the research gap with regard to research software
    engineering (in the human-facing disciplines), where reliability as
    a quality measure has overshadowed validity. The provocative
    question of \emph{to be FAIR or RIGHT} should, of course, be
    answered such, that both are important to ensure sustainable
    research outcomes. Future research should target the third quality
    measure, objectivity, too.

    \printbibliography

\end{document}